# The Widening Gap in Tax Attitudes: Role of Government Trust in the post COVID-19 period

Eiji YAMAMURA

Department of Economics, Seinan Gakuin University/ 6-2-92 Nishijin Sawaraku Fukuoka, 814-8511.

Email: yamaei@seinan-gu.ac.jp

Fumio OHTAKE

Center for Infectious Disease Education and Research, Osaka University, Japan,

Email: ohtake@cider.osaka-u.ac.jp

# Abstract

  This study investigates shifts in acceptable tax rate for reducing inequality during the COVID-19 pandemic using Japanese data. We find a transition from norm-based, unconditional support for redistribution to conditional altruism. Before the pandemic, support remained high and independent of institutional trust. The pandemic generated an overall decline in altruistic attitudes while increasing their dependence on trust in government, particularly among high-income individuals. This "widening gap" implies that in post-crisis societies, the social contract is no longer anchored in stable social norms but increasingly relies on institutional trust to sustain income redistribution from the rich to the poor.

**Keywords:** Altruism, COVID-19, Redistributive attitude, Government Trust

JEL classification. D63; D73; H24; I18;

# 1  Introduction

The pandemic was more than a global health crisis in human history (Alfani, 2022; Saka et al. 2022).  The Spanish flu of 1918 increased in inequality (Galletta and Giommoni, 2022), whereas the fourteenth-century Black Death reduced inequality (Alfani, 2022).  Effects of COVID-19 varied in the present day. Some studies found COVID-19 was associated with increase in inequality (Angelov and Waldenström, 2023; Cortes and Forsythe, 2023), whereas others reported that inequality reduced (Clark et al., 2021).  COVID-19 was considered to serve as a catastrophic stress test for the modern society. Meyer et al. (2025) provided evidence that the COVID-19 pandemic exacerbated long-standing disparities, disproportionately impacting employment, income, and education for vulnerable group.  Many studies explored the short-term macro-level consequences of this shock (e.g., Burlina & Rodriguez-Pose,2024; Crossley et.al. 2023; Furceri et al. 2022). However, long-term impact of COVID 19 on societ has not been sufficiently examined. For instance, its impact on the micro-level "social contract"—specifically, the willingness of citizens to support redistribution—remains a subject should be addressed to maintain society in the post-COVID-19 pandemic period. [1]

A negative shock such as natural disaster would form social capital and increase support for the social safety net (Yamamura 2016 a). However, empirical findings have been paradoxical. Bellani et al. (2023) demonstrated that in Germany, preferences for redistribution actually declined as the crisis intensified, suggesting that collective shocks do not automatically foster solidarity. A crucial mediator in this relationship appears to be institutional trust. As Castro and Martins (2024) argue, during the pandemic, economic confidence and perceptions were shaped less by direct health impacts and more by policy measures and the perceived efficacy of institutional responses. This implies that in times of systemic uncertainty, standard macroeconomic variables play a secondary role to the perceived quality and reliability of the state.

---

[1] Zakharov, and Chapkovski (2025) explored relation between preference for redistribution and war.  Roth and Wohlfart (2018) investigated relation experienced inequality and preferences for redistribution.

The link between trust and fiscal cooperation is well-established in the literature on "tax morale" (Alm and Torgler, 2006, Frey and Torgler, 2007; Kouamé, 2021), which posits that redistributive support is often a form of "conditional cooperation." In the Japanese context, a homogeneous harmonious society is sustained by informal norms based on interpersonal interaction, rather than formal rules (Yamamura 2008). Yamamura (2012) found that social capital and community participation significantly drive redistributive preferences, particularly among high-income groups. Yet, it remains unclear how a crisis like COVID-19 alters this delicate nexus between income capacity, institutional trust, and altruism.

This study used a large-scale individual level dataset from Japan (2018–2025), that were collected through Internet surveys. In the questionnaire, respondents were asked the maximum tax rate they would be willing to accept, given that the tax revenues they pay would be transferred directly to individuals who are significantly poorer than themselves. Using the hypothetical question, we examine how income level and trust in government are associated with individual's acceptable tax rate. Trust in government was found to play a vital role on supporting government's income redistribution policy (Kuziemko et al. 2015, Yamamura 2014a). Different from previous studies, in this paper, we asked individual's own willingness to accept tax rate. We assume that the norm is substitute to trust in government to promote citizen's altruistic behavior by considering their acceptable tax rates. At the onset of COVID-19, short-term analysis has been conducted to explore norm and government response towards emergent situaiton (Hensel et al. 2022; Lazarus et al. 2020). The contribution of this study is to consider structural change by comparing norm and institutional trust covering 8 years before and after COVID -19.

Our findings reveal that prior to the pandemic, support for redistribution in Japan was remarkably resilient, maintained at a high baseline across all income strata regardless of trust in government. However, the COVID-19 shock appears to have eroded these traditional norms. In the post-pandemic era, altruistic attitudes have not only shifted downward but have become sharply bifurcated based on institutional trust.

This "widening gap" in tax attitudes is most acute among high-income earners. While affluent individuals who maintain high trust in government continue to support redistribution, those with low trust have significantly withdrawn their support, effectively acting as "conditional cooperators" who refuse to contribute to a government system they no longer find reliable. This observation aligns with the global trend of political polarization and the fraying of the social contract. By integrating the macro-perspective of inequality (Meyer et al., 2025) with the micro-dynamics of institutional perception (Castro & Martins, 2024), this paper contributes to the literature on public finance and income distribution by demonstrating that institutional trust is the essential prerequisite for mobilizing private resources for public welfare in a post-crisis world.

The remainder of this paper is organised as follows: Section 2 describes the dataset used in this study. Section 3 proposes testable hypotheses and estimation strategies. In Section 4, estimation strategy is explained. Section 5 reports the estimation results and their interpretations. In the final section, implications based on the findings are presented.

# 2 Data

## 2.1 Data Collection

This study used data which has been collected by Internet surveys. We independently and purposefully designed the survey to include hypothetical questions. [2] Before emergence of COVID 19, only a survey has been conducted in 2018, while after spread of the COVID-19, we conducted four times surveys in 2021, 2023, 2024, and 2025, respectively. We used repeated cross-section data.

The survey was administered by Nikkei Research Company (NRC), a professional research firm selected on the basis of its extensive expertise in conducting academic survey research. A structured questionnaire was distributed to eligible participants at the outset of the data collection period. To enhance the representativeness of the sample, the demographic composition of respondents was calibrated to approximate the population structure of Japan at the national level. Accordingly, participants were recruited from across all prefectures, spanning the age range of 20 to 70 years.

---

[2] The primary objective of the survey was to explore broad socio-economic conditions rather than a single specific analysis. Although the survey is comprehensive in nature, it purposefully includes specific items designed to facilitate the analysis conducted in this study.

The total sample size was 18,181, comprising 5,929 respondents surveyed prior to the COVID-19 pandemic and 12,252 surveyed in the post-COVID-19 period. The dependent variable is altruism, operationalized as the acceptable tax rate to be directly transferred to individuals with lower incomes than the respondents. This measure was elicited through the following questions presented under a hypothetical scenario.

> *"Q1. Assume that 80 percent of Japan's population earns less than one-fifth of your income. Further, suppose that the tax paid by you goes directly to those with lower incomes. What percentage of your income would you be willing to pay as tax?*
>
> *Please indicate your allowable tax as above. Choose from 1–50 percent."*

In Q1, the acceptable tax rate takes linear values ranging from 1 to 50 percent, which is employed as a measure of the degree of altruism. This is represented as *Tax Rate*. The approach more directly evaluates individuals' willingness to pay taxes than the widely used Likert-scale measures of redistribution preferences (e.g., Alesina and La Ferrara, 2005; Alesina and Angeletos, 2005; Corneo and Grüner, 2002; Yamamura, 2012, 2014). To keep the situation equal for high- and low-income people in the real world, we assumed that all respondents had a far higher income than average. Thus, we assumed unrealistically large income inequality.

Apart from it, another key variable is trust in government because it is expected to lead individuals to increase redistribution preferences (e.g., Kuziemko et al., 2015; Oh and Hong, 2012; Yamamura, 2014a). To obtain the proxy for trust in government, which is represented as *Trust*, the questionnaire also includes the following question.

Q2: How much do you trust the government? Please select the option that best reflects your view.

1 Do not trust at all, 2 Tend not to trust, 3 Neither trust nor distrust

4 Tend to trust, 5 Trust completely

## 2.2 Basic Statistics

Table 1 provides mean values and standard errors of variables used for estimation in the pre and post COVID-19 periods. In addition, results of their mean difference test were exhibited.

<Table 1>

*Tax Rate* distinctly declined from 13.2 percent to 10.2 percent, which its difference is statistically significant at the 1 % level. Individuals have clearly become less supportive of income redistribution after COVID-19 period than before COVID-19 one. However, both of *Trust* and *Income* (Household income) show almost the same value. In the estimation model shown in the next section, dependent variable (*Tax Rate*) remarkably changed, whereas key independent variable stable (*Tax Rate Trust* and *Income*) before and after COVID-19. Apart from it, standard deviation of *Income* increased from 429 to 659, suggesting an increasing income inequality during the COVID-19 period. This is consistent with previous works (Angelov and Waldenström, 2023; Cortes and Forsythe, 2023).

Interestingly, Married dummy and Female dummy significantly declined, indicating that marriage rate reduced by 5 percent points, and female respondents rate reduced by 4 percent point. This indicates that COVID 19 changed the stable marital relationship, and widening gender gap of motivation to participate in survey.

## 3 Testable Hypotheses

In countries characterised by high trust and low corruption, individuals support high taxation (Algan et al., 2016, 862). However, natural disasters may incentivise bureaucratic corruption (Cevik and Jalles, J.T, 2025; Yamamura, 2014 b). If trust in government erodes, long-term public support for government-led redistribution may decline, potentially resulting in a vicious cycle. In contrast, COVID 19 possibly leads people to higher institutional trust (Esaiasson et al. 2020). Actually, degree of trust in government is stable in the dataset even after experiencing COVID 19 (Table 1). However, marital relationship tended to resolve. One possible interpretation is that the traditional norm to maintain family eroded. This is in line with argument of tradition norm changes to fit the new environment if disastrous events occur (Giuliano and Nunn 2021). Therefore, we assume that COVID-19 impaired the norm to assist the poor for maintaining harmonious society.

To examine the structural transformation from unconditional norms to trust-contingent altruism, we derive three testable hypotheses by integrating the recent literature on pandemic-induced shocks with argument of institutional trust.

3.1. The Erosion of Unconditional Social Norms
Before the pandemic, redistributive preferences in Japan were largely sustained by stable social norms (Yamamura 2008. However, Bellani et al. (2023) found that collective negative shocks can paradoxically reduce support for redistribution in Europe. Furthermore, Meyer et al. (2025) highlight that the pandemic exacerbated existing inequalities, which may have further strained social cohesion. We expect that this shock eroded the "unconditional" nature of altruism.

*Hypothesis 1*: While redistributive support was high and independent of trust levels before the pandemic, the post-pandemic baseline has shifted downward.

3.2. Institutional Trust as a Prerequisite for Conditional Cooperation

The literature on "tax morale" suggests that redistributive support is a form of "conditional cooperation" predicated on the perceived quality of the state (Alm and Torgler, 2006, Frey and Torgler, 2007; Kouamé, 202). Yamamura (2014a) empirically demonstrated a positive correlation between trust in government and preferences for income redistribution in the Japanese context. Given that economic perceptions during the pandemic were dominated by policy efficacy rather than direct health impacts (Castro and Martins, 2024), we hypothesize that trust in government has become the primary determinant for maintaining altruism.

*Hypothesis 2*: Following the pandemic shock, individuals with high government trust maintain high levels of redistributive support, whereas those with low trust significantly withdraw their support.

# 4 Estimation Strategy

**4.1. Empirical Framework of baseline model**

To test the structural shift in willingness to pay tax, we rely on a interaction model that captures the nexus between institutional trust and income level by comparing pre and post COVID-19 period. The baseline specification is as follows:

To explore how household income and trust in government are associated with accepted tax rate, we used the baseline estimation function as shown below:

$$Tax\ Rate_i = \alpha_0 + \alpha_1 Income_i + \alpha_2 Trust_i + \alpha_3 Female\ dummy_i + \alpha_4 Married\ dummy_i + \alpha_5 Age_i + \alpha_6 Age_i^2 + \alpha_7 Schooling\ years + \alpha_8 Schooling\ years_i^2 + e_i$$

First, the dependent variable is *Tax Rate*. In this case, *Allowable Tax* is censored at the lower bound (1 percent level) and upper bound (50 percent level), respectively. Hence, we used a two-limit Tobit model for estimation. The expected sign of *trust* was positive. Control variables are household income, female dummy, age and its square, schooling years and its square. Further, we also control 46 dummy variables to capture prefecture where respondents resided. $e_i$ is an error term. To compare before and after COVID-19 periods, we conducted estimations by using sub-sample of 2018 (pre COVID-19) and 2021, 2023, 2024, and 2025 (post COVID-19). As explained in the previous section, composition of sample differed between before and after the pandemic. To control it, *Married dummy* and *Female dummy*, *Ages* and its square term (*Ages²*) are included. Their results are shown in Table 2.

**4.2. Illustration of results of income and institutional trust**

To facilitate a clear comparison of the structural shifts before and after the pandemic, we present our additional results using coefficient plots in the main text (Figures 1 and 2). Further, we demonstrate visually predicated acceptable tax rates at each institutional trust by comparing different income groups. This approach allows us to focus the discussion on the evolution of the interaction between income and institutional trust which is the primary interest of this study.

In alternative specification of base line model, we consider differences of individual's income positions. *Income* is replaced with four income class dummies, dividing income rankings into quintiles and using the middle-income group as the reference group. Specifically, using the 41–60% income group as the reference category, income position dummies for the 0–20%, 21–40%, 61–80%, and 81–100% income group dummies are included as independent variables. Further, to explore how institutional trust is associated with the acceptable tax, instead of income dummies, we include four dummies of trust in government when reference group is "Neither trust nor distrust". Figure 1 and 2 illustrate coefficients of these dummy variables although results other control variables are not reported because their results are exhibited in Table 2.

To explore the evolving relationship between income position and institutional trust, we employ a full factorial interaction model. We interact the dummy variables for income quintiles and government trust. The middle-income group with neutral government trust serves as the statistical reference group. Value of constant term calculated by Tobit estimation results indicates the level of *Tax Rate* of the reference group after adjusting for other control variables at their means. Coefficients of each interaction term of dummy variables are added to the level of *Tax Rate* of the reference group. Figure 3 (a) and (b) illustrate the predicted value of *Tax Rate* at each institutional trust by showing lines of 41-60% income and 81-100%. This approach allows us to observe the actual predicted levels of tax-acceptability across groups of institutional trust and income, rather than mere relative deviations from a baseline group although we focus on the contrast between high and middle income groups.

## 5 Estimation results and its interpretation

### 5.1 Baseline results

Table 2 shows negative and positive signs of coefficient of *Income* in pre and post COVID-19 periods, respectively. The statistical significance is observed only in post COVID-19 period, implying that individuals in higher income household come to accept higher tax rate. One possible interpretation is, before the pandemic, a shared social norm regarding the obligation to pay taxes was prevalent, resulting in high tax morale across all income levels. However, this sense of normative duty has weakened in the post-pandemic period.

<Table 2>

Turning to *Trust,* its coefficients show the predicted positive sign and statistical significance at the 1 % level in both periods. This is linear effect of *Trust.* To scrutinize it, we used category dummies of trust in government and its results are reported in Figures 2 and 3 in the following section.

### 5.2 Comparing between different groups regarding income and institutional trust.

### 5.2.1. Income Quintiles and Altruism

Figure 1 illustrates the coefficients of income quintiles relative to the middle-income group (41–60%). While altruistic differences across income levels were modest in 2019 (Pre-COVID, blue dashed lines), a divergence emerged in 2020 (Post-COVID, red solid lines). Specifically, the top income quintile (81–100%) showed a sharp and statistically significant increase in acceptable tax rate compared to the pre-pandemic period. Relation between income level and the acceptable tax rate is positive but not linear. We interpreted it as follows. The pandemic exacerbated income inequality (Table 1). Consequently, growing resentment among the poor toward high-income earners has led to social instability. As a result, high-income individuals have become more sensitive to criticism from the impoverished and more conscious of the significance of narrowing the gap between the rich and the poor (Yamamura 2016b).

<Figure 1>

### 5.2.2 Government Trust and Altruism

Figure 2 shows the moderating effect of institutional trust. Relative to the neutral group, as a whole, high and low trust groups show high and low acceptable tax rate, respectively. However, before COVID 19, we should draw attention to no significant difference of the acceptable tax rate between the highest trust and neutral groups. In contrast, individuals with high government trust (levels 4 and 5) exhibited a marked surge in acceptable tax rate in post-pandemic period. This suggests that government trust acts as a crucial catalyst for prosocial behavior during period when individuals confronted social instability and increase in income inequality.

<Figure 2>

All in all, what is observed in Table 2, Figures 1 and 2 support *Hypotheses 1 and 2.* This clearly indicates change of individual's accepted tax rates and so altruistic redistributive motivation.

### 5.3 Structural Changes: Interpretation of Figure 3

For closer examination, we examined the interaction between income and trust in government. The core of our empirical contribution is visualized in **Figure 3**, which presents the predictive acceptable tax rate, which is regarded as redistributive altruism across levels of trust and income. To avoid any misinterpretation of the results, it is essential to focus on the highest and mid-income groups to clearly show the major findings.

<Figure 3(a)>

<Figure 3(b)>

In the pre COVID-19 period, Figure 3(a), the lines for both income groups are relatively flat. This indicates that redistributive support was based on an "unconditional social norm". In this period, individuals supported the social safety net regardless of their trust in the government. In contrast, in Figure 3(b), we observe the positive slope for both median and high income group. This upward trajectory confirms that altruism has transitioned into a **"conditional altruism"** predicated on institutional trust (**Frey & Torgler, 2007**).

Considering Figure 3 (a) and (b) indicates a general shift toward conditional cooperation during the pandemic period. In addition, Figure 3(b) demonstrate steeper slope for the high-income group (81-100%) highlights an intensified 'trust-dependency' in their redistributive preferences, leading to a widening gap in social solidarity at higher levels of institutional trust. While the 95% confidence intervals overlap, indicating that the gap between strata is not always statistically distinct at every point, the overall shift from a fragmented pre-pandemic distribution to trust-dependent consensus in the post-period is evident. The increase in accepted tax was most pronounced among high-income individuals who also maintained high trust in the government.

Apart from it, in Figure 3 (a), level of the accepted tax rate is around 11 % at the lowest trust group regardless of income group. The tax rate is almost the same for high income group while the rate slightly increases to 12 % for middle income group. After entering the COVID-19 period, surprisingly, level of the accepted tax rate is 6 % at the lowest trust group for income group for both income groups. So, the accepted tax rate declined by 5 % from the pre-pandemic period. However, the rate at the highest trust group is 12% and 9 % for highest and middle income groups, respectively. Interestingly, in case of the lowest income group, the acceptable tax rate is lower after the COVID-19 than before it. However, in case of the highest income group, the rate is higher after the COVID-19 than before it while being only observed for high income group. Concerning the gap of acceptable tax rate for high trust group, the rich people accepted lower tax rate by 1 % before the pandemic, but higher tax rate by 3% after the pandemic.

Comparing the two periods reveals a downward shift in the overall level of altruism at lower trust levels. Consistent with **Bellani et al. (2023)**, the pandemic shock eroded the baseline of solidarity. Post-COVID, altruistic behavior is no longer a default setting. Following the logic of **Yamamura (2012, 2014, 2016)**, high-income earners possess greater economic capacity and are more sensitive to psychological externalities and institutional quality. The gap between high-trust and low-trust individuals has dramatically widened for the high income group. This **"Widening Gap"** demonstrates that while the affluent can be mobilized for redistribution through high institutional trust, they are also the quickest to withdraw support when trust is compromised.

**5.4 Discussion**

Different from previous works to use Likert scale of preference for redistribution((e.g., Alesina and La Ferrara, 2005; Alesina and Angeletos, 2005; Corneo and Grüner, 2002; Yamamura, 2012, 2014)), we use the accepted pay the tax rate exclusively transferred to the poorer individuals than respondents. The contribution of this study is to explore how the altruistic income redistribution has been changed during the COVID-19 pandemic by independently and purposefully collecting the individua's accepted tax by hypothetical questions though internet survey.

Our analysis reveals a fundamental shift in the motivators of income redistribution. While overall support for redistribution became more averse to redistribution, it became highly sensitive to the level of government trust. Specifically, for high-income earners, altruism is no longer a given; it is now strictly contingent upon their trust in government. This implies that in a post-crisis environment, trust is critical to mobilize private resources for public welfare.

The COVID-19 pandemic triggered a divergence, where economic resources and institutional trust became positively associated with the accepted tax rates. These findings imply that policy effectiveness to reduce income inequality depends heavily on trust in government. Both institutional trust play a more vital role to sustain or enhance social stability even after the unconditional norm has eroded.

# 6 Conclusion

This study reveals a structural shift in the factors associated with acceptable tax to be transferred to the poor, transitioning from unconditional social norms to trust-contingent altruism in the context of the COVID-19 pandemic. Prior to the crisis, redistributive support in Japan was consistently high and independent of institutional trust, suggesting that altruism was anchored in stable, pre-existing social norms. These norms were sustained by social capital—the interpersonal trust and cooperative dispositions that accumulate through repeated face-to-face interactions among community members in harmonious Japanese society (Yamamura 2008).

However, the pandemic period is associated with an erosion of these norms, accompanied by an overall decline in altruistic attitudes and a stronger dependence on government trust—a pattern particularly pronounced among high-income earners. One plausible interpretation, though not directly tested in this study, is that the pandemic-induced reduction in face-to-face interaction depleted the social capital that had previously underpinned redistributive norms, thereby weakening the interpersonal foundations of unconditional altruism. This remains an implication drawn from our empirical findings rather than a statistically verified claim.

Our findings are highly consistent with the global trend of increasing political polarization, social fragmentation, and the deterioration of established social norms. As crisis-induced uncertainty deepens, the "social bond" that once bound disparate socioeconomic groups together through unconditional solidarity appears to be giving way to a more calculated, conditional cooperation. This "widening gap" in tax attitudes poses a significant challenge to the sustainability of the social contract; where institutional trust remains low, even the affluent may withdraw their support for redistribution precisely when social inequality is most acute.

While this study provides evidence of this structural shift, several questions remain. Future research should investigate whether this reliance on trust reflects a permanent transformation of social preferences or a temporary defensive response to a major shock. Furthermore, it is crucial to explore how specific dimensions of government performance—such as the transparency of wealth redistribution and the effectiveness of social safety nets—may revitalize altruism in an increasingly polarized world. Addressing these questions, which are vital for rebuilding inclusive welfare systems to withstand future systemic shocks.

Journal of Public Economics, 167, 251–262.

Saka, O., Eichengreen, B., Aksoy, C. (2022). The political scar of epidemics. Economic Journal, 134(660), 1683–1700.

Yamamura, E. (2008) The market for lawyers and social capital: Are informal rules a substitute for formal ones? Review of Law & Economics, 4 (1), 499-517.

Yamamura E. (2012). Social capital household income and preferences for income redistribution. European Journal of Political Economy 28(4) 498-511.

Yamamura, E. (2014a). Trust in government and its effect on preferences for income redistribution and perceived tax burden. Economics of Governance 15(1) 71-100.

Yamamura,E. 2014b. "Impact of natural disaster on public sector corruption" *Public Choice*, 61(3-4), 385-405.

Yamamura, E. (2016a). Natural disasters and social capital formation: The impact of the Great Hanshin-Awaji earthquake. Papers in Regional Science 95(S1), 143-164.

Yamamura, E. (2016 b). Social conflict and redistributive preferences among the rich and the poor: testing the hypothesis of Acemoglu and Robinson. Journal of Applied Economics, 19(1), 41-64.

Zakharov, A., Chapkovski, P. (2025). The effect of war on redistribution preferences. Journal of Public Economics, 241(C).
23

Table 1. Summary of mean difference test by COVID Groups

| Variable | Pre-COVID19 Mean (SD) | Post-COVID19 Mean (SD) | Mean Difference, p-value |
|---|---|---|---|
| *Tax rate* | 13.22 (12.01) | 10.24 (11.27) | 2.99, p = 0.000 *** |
| *Trust* | 2.53 (1.08) | 2.50 (1.09) | 0.03, p = 0.056 |
| *Income* | 654.90 (429.72) | 659.01 (462.56) | -4.11, p = 0.565 |
| *Schooling years* | 14.72 (1.95) | 14.76 (2.00) | -0.05, p = 0.095 |
| *Married dummy* | 0.55 (0.50) | 0.51 (0.50) | 0.04, p = 0.000 *** |
| *Age* | 44.76 (12.63) | 46.88 (13.89) | -2.11, p = 0.000 *** |
| *Female dummy* | 0.49 (0.50) | 0.45 (0.50) | 0.04, p = 0.000 *** |
| Observation | 5,929 | 12,252 | |

Note: Pre and Post COVID-19 refer to the two comparison groups. Asterisks indicate statistical significance: *p < 0.01, p < 0.05. Sample size is the same as used in Table 2.

Table 2. Tobit Regression Results: Dependent variable is *Tax rate*

| Variable | Pre-COVID19 (1) | Post-COVID19 (2) |
|---|---|---|
| *Income* | -0.063 (0.453) | 1.406 (0.318) *** |
| *Trust* | 0.682 (0.168) *** | 1.421 (0.123) *** |
| *Female dummy* | -3.513 (0.369) *** | -2.520 (0.276) *** |
| *Married dummy* | -0.196 (0.412) | -0.425 (0.307) |
| *Age* | -0.240 (0.107) ** | -0.359 (0.069) *** |
| *Age$^2$* | 0.004 (0.001) *** | 0.005 (0.001) *** |
| *Schooling years* | 0.225 (1.199) | -2.718 (0.866) *** |
| *Schooling years$^2$* | 0.010 (0.042) | 0.112 (0.030) *** |
| Total observations | 5,929 | 12,252 |
| Uncensored | 5,039 | 8,991 |
| Left-censored | 669 | 2,905 |
| Right-censored | 221 | 356 |
| LR chi² | 252.17 | 517.03 |
| Prob > chi² | 0.0000 | 0.0000 |

Note: Coefficients are followed by standard errors in parentheses. Significance levels: *p < 0.01, p < 0.05, *p < 0.10*. Income coefficients and SEs are scaled by 1000 for interpretability. Results of residential prefecture dummies are excluded from this table.

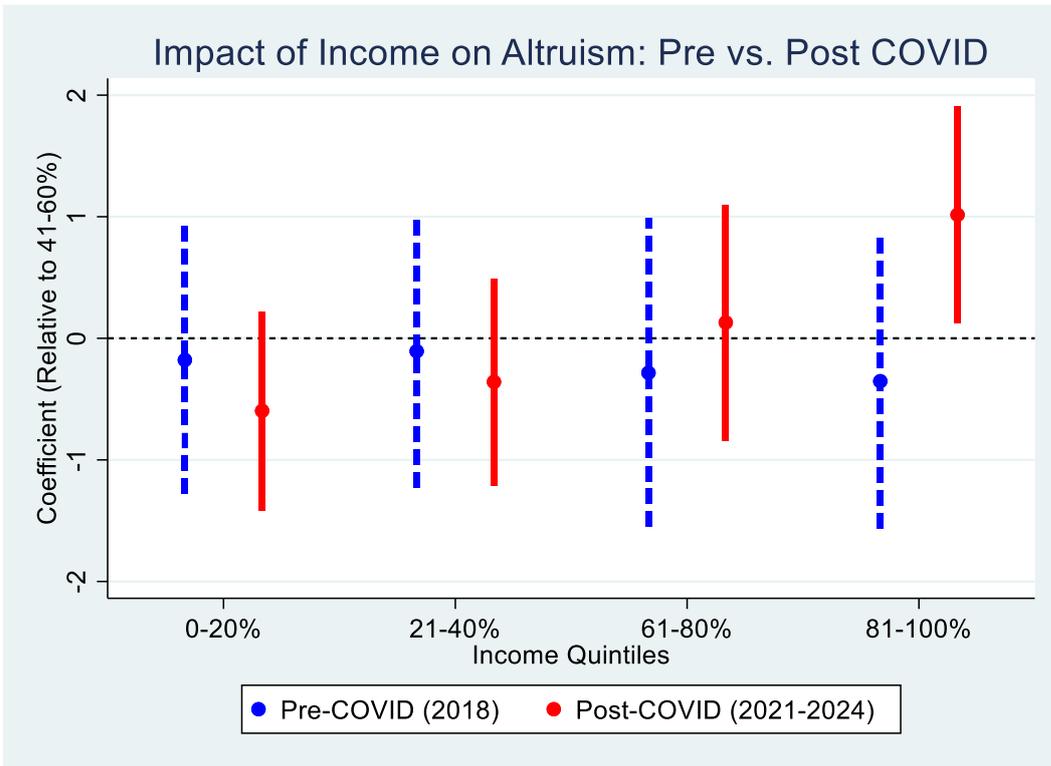

**Figure 1:** Impact of Income on Altruism: Pre vs. Post COVID (Tobit)

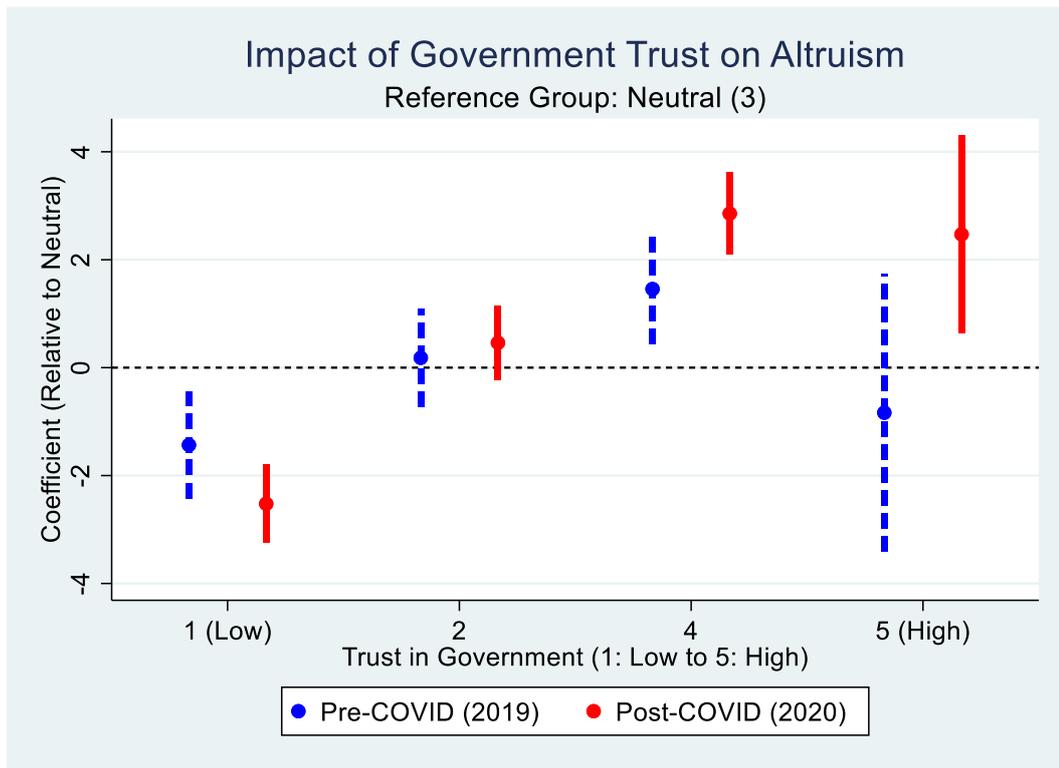

**Figure 2: Impact of Government Trust on Altruism**

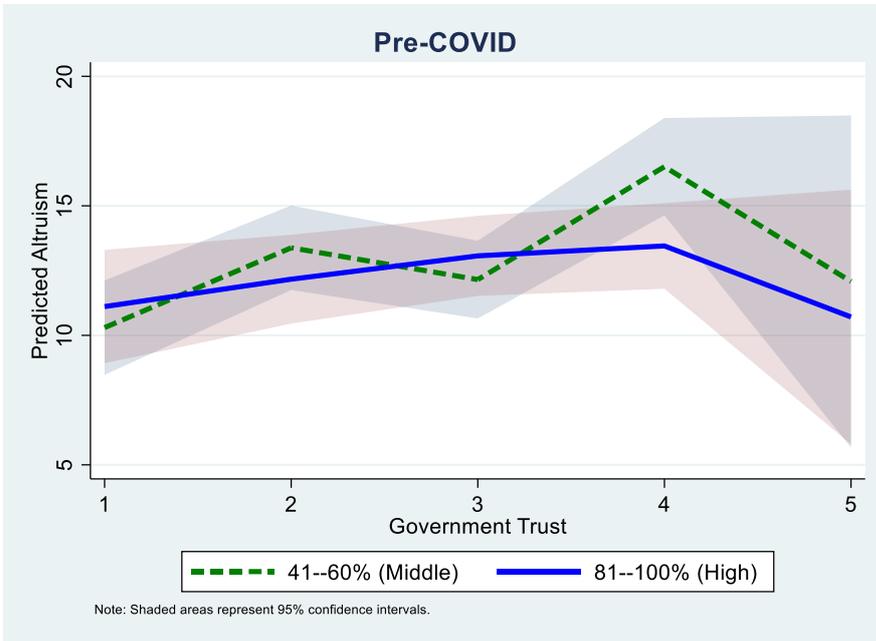

**(a) Pre-Covid period (2018)**

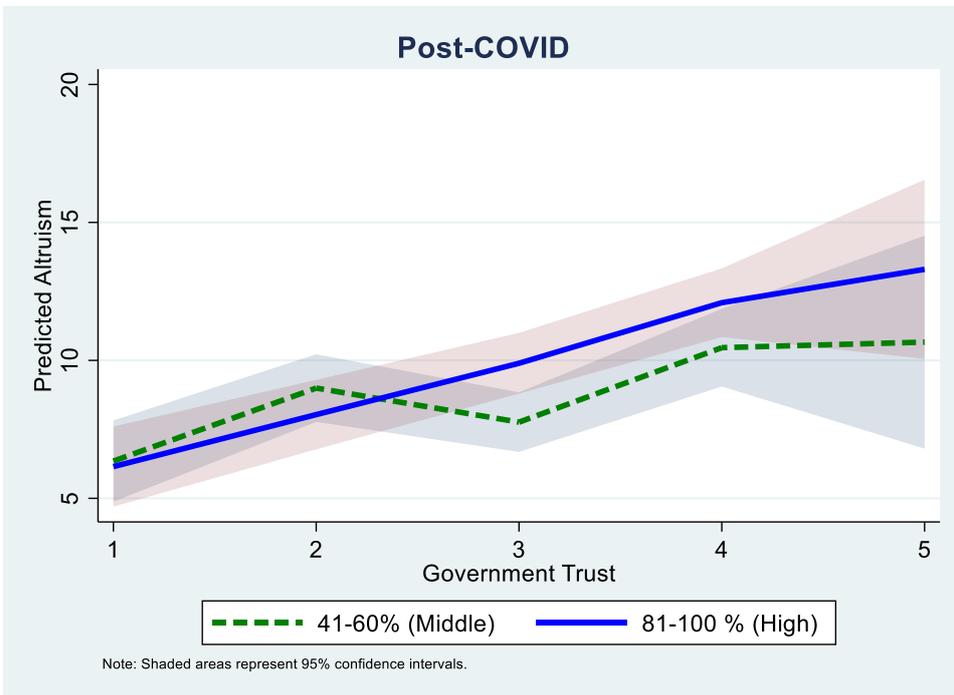

**(b) Post-Covid period (2021-2025)**

**Figure 3: Predicted Accepted Tax by Income and Trust (Post-COVID).**